\documentclass[12pt]{iopart}
\usepackage{iopams}  
\usepackage{harvard}
\usepackage{graphicx}
\usepackage{aas_macros}

\newcommand{\msun}{\mbox{M}_\odot}

\begin{document}

\title[]{Double white dwarfs and \emph{LISA}}

\author{T.R. Marsh}

\address{Department of Physics, University of Warwick, Gibbet Hill Road,
Coventry CV4 7AL}

\ead{t.r.marsh@warwick.ac.uk}
\begin{abstract}
Close pairs of white dwarfs are potential progenitors of Type~Ia supernovae
and they are common, with of order 100 -- 300 million in the Galaxy. As such
they will be significant, probably dominant, sources of the gravitational waves
detectable by \emph{LISA}. In the context of \emph{LISA}'s goals for
fundamental physics, double white dwarfs are a source of noise, but from an
astrophysical perspective, they are of considerable interest in their own
right. In this paper I discuss our current knowledge of double white dwarfs
and their close relatives (and possible descendants) the AM~CVn
stars. \emph{LISA} will add to our knowledge of these systems by providing the
following unique constraints: (i) an almost direct measurement of the Galactic
merger rate of DWDs from the detection of short period systems and their
period evolution, (ii) an accurate and precise normalisation of binary
evolution models at the shortest periods, (iii) a determination of the
evolutionary pathways to the formation of AM~CVn stars, (iv) measurements
of the influence of tidal coupling in white dwarfs and its significance for
stabilising mass transfer, and (v) discovery of numerous examples of eclipsing
white dwarfs with the potential for optical follow-up to test models of white
dwarfs.
\end{abstract}


\submitto{\CQG}
\maketitle

\section{Introduction}
In the early 1980s it was suggested that Type~Ia supernovae might come from
close pairs of white dwarfs merging under the action gravitational radiation
losses \cite{Webbink1984,IbenTutukov1984}. It was later realised that the
large number of systems needed to sustain the Type~Ia rate within the Galaxy
under these models meant that double white dwarfs (henceforth DWDs) are likely
to be a dominant source of gravitational waves for space-based interferometry,
to the extent that over some frequency intervals of interest in the context of
\emph{LISA}, DWDs may define \emph{LISA}'s noise floor \cite{Evans1987,Hils1990}.

Early searches for DWDs produced meagre returns, and predictions that
10\% of all ``single'' white dwarfs might in fact be double
\cite{Paczynski1985} seemed wide of the mark. \citeasnoun{Robinson1987} found
no DWDs amongst 44 targets, \citeasnoun{Foss1991} none amongst 25, and 
\citeasnoun{Bragaglia1990} found one certain DWD together with a
few candidates from 54 targets. Together with the system L870-2, discovered by
\citeasnoun{Saffer1988}, by the early 1990s only two DWDs had measured
periods. This changed when advances in our understanding of white dwarf
atmospheres led to the identification of white dwarfs of too low a mass for
single star evolution \cite{Bergeron1992}. Optical spectroscopy
showed that a large fraction of these objects are DWDs
\cite{Marsh:WD1101+364,Marsh:friends,Holberg:Feige55,Moran:WD0957-666,Maxted:WD1704+481}. Since
the turn of the millennium further discoveries have followed from the SPY
survey \cite{Napiwotzki:SPY2} and from the SDSS as detailed later.

These discoveries have established the presence of large numbers of
DWDs and their importance as gravitational wave sources. There have been
numerous studies of the likely impact of DWDs on \emph{LISA}. These find that
below a cutoff frequency of about $2$ to $6\,$mHz, there are
so many systems that their signals are unresolved, while above
this frequency individual systems are resolved, with the odd
nearby system rising above the noise at somewhat lower frequencies
\cite{Hils1990,Hils2000,Nelemans2001,Ruiter2010,Liu2010,Yu2010}.

Most of these studies have been concerned with predicting the gravitational
wave (GW) signal from DWDs in \emph{LISA}. My interest here is more what we can
learn about DWDs from \emph{LISA} that is hard to deduce from 
electromagnetic (EM) observations. The potential is great, with direct
measurement of tidal coupling between white dwarfs and the first detections of
DWDs in globular clusters where their numbers are expected to be dynamically
enhanced \cite{Shara2002}, likely to come from \emph{LISA} data.

\section{The two types of double white dwarfs}
DWDs split into two groups which have different properties, from both the EM
and GW perspectives, although in each case the two groups, while physically
distinct, can be difficult to distinguish observationally. The two groups are
the detached and semi-detached DWDs.  Detached DWDs are simple pairs of white
dwarfs evolving towards shorter periods under the action of gravitational wave
losses. The semi-detached systems, observationally identified as the AM~CVn
stars (see \citeasnoun{Solheim2010} for a recent review), are systems in which
stable mass transfer takes place from a Roche-lobe filling hydrogen-deficient
star to a more massive companion white dwarf. From now on I will refer to all
such systems as AM~CVn stars. Hydrogen deficiency is
necessary to reach short orbital periods; the hydrogen-rich counterparts to
AM~CVn stars are the cataclysmic variable stars which reach a minimum orbital
period of around 80 minutes; I do not consider these further here. The
Roche-lobe filling stars in the AM~CVn stars must be at least partially
degenerate to reach very short orbital periods. White dwarfs fit the bill, and
although these systems are not necessarily ``double white dwarfs'' for
simplicity I will continue to use the umbrella term ``DWDs'' for both classes.

The key difference between these two classes from an EM point of view is the
presence of accretion in the AM~CVn stars which can produce X-rays, atomic
line emission and photometric variability. This is both a blessing and a
curse: a blessing as it makes these systems, which are rare, easier to find,
and a curse because we don't understand accretion well enough to estimate
selection effects with certainty. From a GW standpoint, the key differences
are (a) the system masses which in the case of the AM~CVn stars can reach very
low values ($< 0.1 \,\msun$) inaccessible to the detached DWDs, and (b) the
time derivatives of the gravitational wave frequencies which on the whole will
be negative for the AM~CVn stars but always positive for the detached DWDs.

\section{Detached DWDs}
For \emph{LISA}, detached DWDs (for this section I will drop the
``detached'' qualifier) will probably be the single dominant source class, the
``main sequence'' of space-based gravitational wave astronomy. They are a
class of huge current interest as they are the candidate progenitors of
Type~Ia supernovae usually referred to as the ``double degenerate'' (DD)
scenario \cite{Webbink1984,IbenTutukov1984}. The fortunes of DWDs as Type~Ia
progenitors have waxed and waned over the years when squared up against the
``single degenerate'' (SD) model which supposes accretion from a hydrogen rich
companion \cite{Whelan1973,Nomoto1982}. Recent papers continue to show a lack
of consensus \cite{Gilfanov2010,DiStefano2010} and it is of course possible
that there are multiple progenitor classes, as suggested by evidence for
bimodality in the delay time distribution of Type~Ias \cite{Mannucci2006,Ruiter2009}.

The early failures to find many DWDs have often been raised to argue against
DDs as potential Type~Ia supernova progenitors \cite{Branch1995,Hachisu1999},
indeed, this perception remains current \cite{Parthasarathy2007}. My view is
that, within the admittedly rather large margins of error, this is not a huge
problem given that DWDs with short merger times and ones with total masses
close to the Chandrasekhar limit have been discovered
\cite{Moran:WD0957-666,Napiwotzki2002,Karl2003}. It is perhaps not often
realised that the current sample of DWDs remains strongly biassed towards low
mass systems because these were specifically targeted in the searches that
started in the 1990s as well as in more recent searches. Similarly, the
enormous difference in our ability to find DD versus SD progenitors should not
be underestimated: while it is possible to see potential SD Type~Ia
progenitors in other galaxies (although not necessarily to recognise them as
such), it is hard to follow DWDs using EM observations if they are more than a
few hundred parsecs away: finding DWDs is hard work. The best prospect for an
observational calibration of DWD numbers is offered by the SPY survey
\cite{Napiwotzki:SPY2} that did not target particular mass ranges, although
even it suffers unavoidable selection biases with respect to both mass and
temperature that need allowing for.

To understand what \emph{LISA} can bring to the study of DWDs, it is important
to know first what EM observations can tell us. Table~\ref{tab:detached}
\begin{table}
\caption{Detached double white dwarfs ordered by
  orbital period. The references are to the discovery papers. Objects starting with
  'J' are SDSS white dwarfs. \label{tab:detached}}

\begin{tabular}{lcccllcccl}
\br
Name       & $P$     & $M_1$   & $M_2$    & Rf &
Name       & $P$     & $M_1$   & $M_2$    & Rf \\
           & days    & $\msun$ & $\msun$ & &
           & days    & $\msun$ & $\msun$ & \\
\mr 
J1053+5200 & $0.043$ & $0.20$  & $>0.26$ & 1 & 
PG1713+332 & $1.127$ & $0.35$  & $>0.18$ & 5\\
J1436+5010 & $0.046$ & $0.24$  & $>0.46$ & 1 & 
WD1428+373 & $1.157$ & $0.35$  & $>0.23$ & 15\\
WD0957-666 & $0.061$ & $0.37$  & $0.32$  & 2 & 
WD1022+050 & $1.157$ & $0.39$  & $>0.28$ & 15\\
J0849+0445 & $0.079$ & $0.17$  & $>0.64$ & 3 &
WD0136+768 & $1.407$  & $0.47$  & $0.37$  & 16\\
WD1704+481 & $0.145$ & $0.39$  & $0.56$  & 4 & 
WD1202+608 & $1.493$ & $0.3$   & $>0.25$ & 17\\
PG1101+364 & $0.145$ & $0.36$  & $0.31$  & 5 & 
WD0135-052 & $1.556$ & $0.47$  & $0.52$  & 18\\
PG2331+290 & $0.166$ & $0.39$  & $>0.32$  & 6 &
WD1204+450 & $1.603$ & $0.46$  & $0.52$  & 16\\
J1257+5428 & $0.190$ & $0.20$  & $>0.95$ & 7, 8 &
WD0326-273 & $1.875$ & $0.51$  & $>0.59$ & 12\\
NLTT~11748  & $0.236$ & $0.15$  & $0.71$  & 9 & 
WD1349+144 & $2.209$ & $0.44$  & $0.44$  & 19\\
J0822+2753 & $0.244$ & $0.17$  & $>0.76$ & 3 & 
HE1511-0448& $3.222$ & $0.48$  & $>0.46$ & 12\\
HE2209-1444& $0.277$ & $0.58$  & $0.58$  & 10 & 
PG1241-010 & $3.347$ & $0.31$  & $>0.37$ & 6 \\
J0917+4638 & $0.316$ & $0.17$  & $>0.28$ & 11 & 
PG1317+453 & $4.872$ & $0.33$  & $>0.42$ & 6 \\
WD1013-010 & $0.437$ & $0.44$  & $>0.38$ & 12 &
WD2032+188 & $5.085$ & $0.41$  & $>0.47$ & 6 \\ 
HE1414-0848& $0.518$ & $0.71$  & $0.52$  & 13 &
WD1824+040 & $6.266$ & $0.43$  & $>0.52$ & 15\\
WD1210+140 & $0.642$ & $0.23$  & $>0.38$ & 12 &
WD1117+166 & $30.09$ & $0.7$   & $0.7$   & 20\\
LP 400-22  & $1.010$ & $0.19$  & $>0.41$ & 14 &
           &         &         &         &\\
\br

\end{tabular}
{\tiny
1. \citeasnoun{Mullally2009}, 
2. \citeasnoun{Moran:WD0957-666}, 
3. \citeasnoun{Kilic2010},
4. \citeasnoun{Maxted:WD1704+481}, 
5. \citeasnoun{Marsh:WD1101+364},
6. \citeasnoun{Marsh:friends}, 
7. \citeasnoun{Badenes:1257},
8. \citeasnoun{Kulkarni:1257},
9. \citeasnoun{Steinfadt:NLTT},
10. \citeasnoun{Karl2003},
11. \citeasnoun{Kilic2007},
12. \citeasnoun{Nelemans2005},
13, \citeasnoun{Napiwotzki2002},
14. \citeasnoun{Kilic2009},
15. \citeasnoun{Morales2005},
16. \citeasnoun{Maxted2002},
17. \citeasnoun{Holberg:Feige55},
18. \citeasnoun{Saffer1988}, 
19. \citeasnoun{Karl2003b},
20. \citeasnoun{Maxted:PG1115}.
}

\end{table}
lists the periods and masses of DWDs with published orbital periods. The mass
of the brighter component can usually be measured by modelling its optical
spectrum. Sometimes both components are visible and then both masses can be
measured, but often one can only deduce a lower limit to the mass of the
unseen component from the orbital motion of its companion.  One can sometimes
measure the temperatures of both components and thus the difference between the
formation times of each component, a strong discriminator of the prior
evolution \cite{vanderSluys2006}. The number of detached DWDs in the Galaxy
can approximately be estimated from the fraction of systems observed to be DWD
and the total number of white dwarfs in the Galaxy. This approach gives a
number of systems ranging from 20 to 200 million
\cite{Maxted1999,Holberg2008}. Binary population synthesis studies have given
numbers from around 100 to 400 million
\cite{Han1998,Nelemans2001,Liu2010,Yu2010}.

In comparison with the best EM observations, \emph{LISA} will give us
comparatively limited information on individual systems, yet there are several
ways in which \emph{LISA} can provide greatly superior information on DWDs as
a whole, as I now discuss.

\subsection{Population statistics}
At high enough frequencies, \emph{LISA} will be sensitive to DWDs throughout the
Galaxy and will give us a view of the whole population with relatively little
selection. Several studies have predicted that $\sim 10$,$000$ DWDs should
have high enough frequencies to be resolvable by \emph{LISA}
\cite{Nelemans2001,Ruiter2010}. These will be the shortest period systems,
which are those of most relevance to the merger rate of DWDs, a quantity of
great interest in the context of Type~Ia supernovae. EM observations,
which are only sensitive out to a limited distance, will always be handicapped
in comparison.  Assuming a single chirp mass, $M_c = M_1^{3/5} M_2^{3/5} (M_1+M_2)^{-1/5}$,
the flux of DWDs crossing orbital period $P$ at a time $t_0$ since the formation of the Galaxy is given
by
\begin{equation}
F(P,t_0) = - n(P,t_0) \dot{P} = \int_P^{P_{m}} B(P',t_0 - \tau(P' \rightarrow P))
\, dP' , \label{eq:flux}
\end{equation}
where $\tau(P' \rightarrow P)$ is the time taken for a system to change period
from $P'$ to $P$, $n(P,t)$ is the orbital period distribution and $B(P,t)$ is
the birth rate period distribution at time $t$. A more realistic model would
require integration over the distribution of chirp masses as well. The upper period
limit $P_{m}$ is set by the maximum period that is able to evolve to period
$P$ within the lifetime of the galaxy, i.e.
\begin{equation}
\tau(P_m \rightarrow P) = t_0.
\end{equation}
For the short period systems that we are interested in for \emph{LISA}, $P_{m}$
is around $5$ to $15\,$hours. As we approach short periods ($P \rightarrow 0$)
$P_{m}$ will tend to a constant and thus the flux $F = -n \dot{P}$ will tend to a
constant, i.e. the DWD merger rate. The numbers per unit period then scale as
$n \propto 1/\dot{P}$, or equivalently $n \propto \tau_m/P \propto P^{5/3}$ 
where $\tau_m$ is the merger time at period $P$
\begin{equation}
\tau_m = 1.00 \times 10^7 (M_c/\msun)^{-5/3} (P/1\,\mbox{h})^{8/3} \, \mbox{yr},
\end{equation}
assuming that we can neglect the effect of tides, although these are likely to
be significant at these short periods \cite{Willems2007}.
The rapid reduction in lifetime with period makes it hard for EM-based
searches to probe the short-period end of the birth rate distribution
directly; this was realised by \citeasnoun{Robinson1987} as the major caveat
on their null result. Looking at Table~\ref{tab:detached}, EM observations are
unlikely to provide strong constraints upon the integrand of Eq.~\ref{eq:flux}
for periods much below one hour. At such periods the merger times are of order
10 to 100 million years, depending upon mass, i.e. at least 100 times shorter
than the age of the Galaxy, and still 20 times shorter than the time for which
white dwarfs display strong spectral features. It is probably no coincidence
that the shortest period systems known have low masses since this increases
their survival time.

\subsection{DWD sub-types}
\begin{figure}
\centering
\includegraphics[angle=270,width=0.5\textwidth]{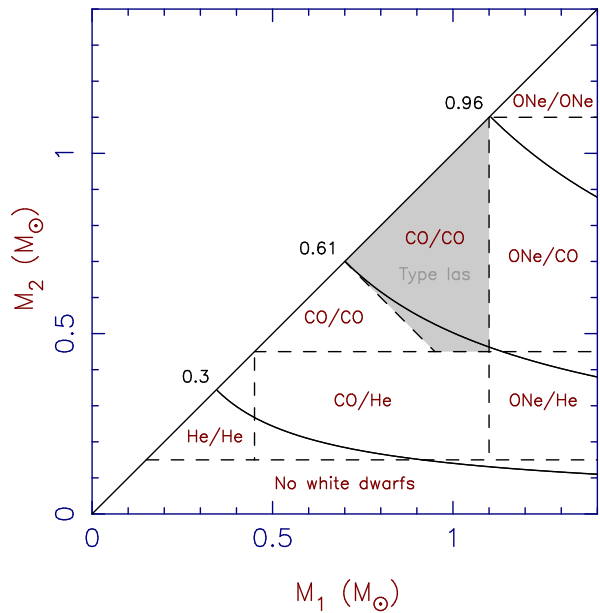}
\caption{\label{fig:pspace} The curved lines are lines of constant chirp mass;
the straight regions delineate different pairs of DWDs, assuming a unique
mapping between bulk composition and mass: He $= 0.15$ -- $0.45$, CO $= 0.45$
-- $1.1$, ONe $= 1.1$ -- $1.4\,\msun$.}
\end{figure}

For some fraction of the resolved \emph{LISA} sources, it will be possible to
detect not just the frequency $f$, but its time derivative $\dot{f}$. For
detached DWDs this is a function of period and chirp mass only. The best EM
observations can return $M_1$ and $M_2$ separately, but \emph{LISA} wins
through the very large number of likely detections, sensitivity to those of
high mass and short period, and well-understood selection effects. The outcome
of DWD mergers depends upon their masses and their composition. For instance
the canonical Type~Ia model for DWDs involves the merger of two carbon-oxygen
white dwarfs. To a large extent, the bulk composition of white
dwarfs is thought to map into their mass. If so then, as Fig.~\ref{fig:pspace}
illustrates, chirp mass measurements have good potential when combined with
population synthesis to discriminate the fraction of potential Type~Ia
supernovae, pairs of helium white dwarfs, etc. Example numbers are around
$10$,$000$ resolvable DWDs \emph{LISA} \cite{Nelemans2003,Ruiter2010,Liu2010},
around 600 of which will have detectable frequency changes within
one year \cite{Nelemans2003,Ruiter2010}, and presumably many more over longer
intervals. The different types (He+He etc) lead to a very strongly structured chirp
mass distribution illustrated in Fig.~7 of \citeasnoun{Liu2010} supporting the
potential that chirp masses hold for probe DWD evolution.

\section{AM~CVn stars}
\begin{figure}
\centering
\includegraphics[angle=270,width=0.5\textwidth]{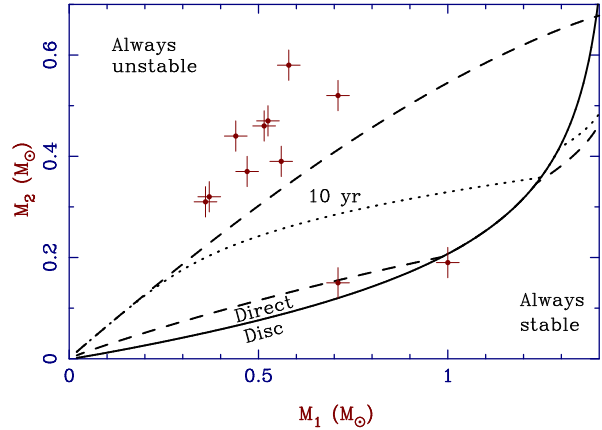}
\caption{\label{fig:merge} Stability regions for mass transfer between two
  white dwarfs (from star 2 to star 1), based on
  \protect\citeasnoun{Marsh:2004} with systems of known mass from
  Table~\protect\ref{tab:detached} over-plotted. The upper dashed line marks the
  dividing line between stability and instability if the accretor does not
  gain angular momentum, while the lower solid line applies if it is not
  coupled to the orbit. Until the recent discoveries of the two systems lowest
  on the plot, all known systems were unconditionally unstable.
}
\end{figure}

In an AM~CVn star, degenerate or semi-degenerate donor stars lose mass to
white dwarf companions (note that there are similar systems, the ultra-compact
X-ray binaries in which the accretors are neutron stars). As they do so, they
expand, and the orbital periods lengthen. \emph{LISA} sources with $\dot{f} <
0$ are therefore likely AM~CVn stars.  The known systems have orbital periods
that range from just over 5 minutes to 65 minutes. Periods this short require
the mass donors to be largely or entirely hydrogen-deficient in order to be
dense enough to fit within their Roche lobes. Most known examples do indeed
lack hydrogen, the exception being HM~Cnc \cite{Reinsch2007}.  A key unsolved
issue for these systems is how they form. Attention has focussed on three
types of progenitor: (i) detached DWDs, (ii) evolved cataclysmic variable
stars, and (iii) white dwarf / helium star accreting binaries. At the long
periods of most known systems, these three models lead to rather subtle
differences in mass transfer rate and other parameters \cite{Deloye:2007}
which even the best constrained systems are not yet capable of distinguishing
\cite{Copperwheat:2010}. We know DWDs of short enough period to merge within a
Hubble time, while there are no clear progenitors of the other two
routes. However, it is not obvious that DWDs will survive the onset of mass
transfer because of the instability that can set in if the two white dwarfs
are of a similar mass \cite{Marsh:2004}. Indeed, until recently all DWDs of
known mass ratio were candidates for merging (Fig.~\ref{fig:merge}).

The differences are more marked at short periods. For instance, only DWDs are
thought to be able to reach periods well below 10 minutes. HM~Cnc ($P
= 321\,$sec.) is the only such system known \cite{Ramsay:2002,Roelofs:2010}.
HM~Cnc is optically faint and has a soft X-ray spectrum that could easily be
absorbed if it were more distant. \emph{LISA}'s sensitivity to short period
systems holds great potential for finding more such systems and for
elucidating the nature of their donors. This can first be carried out in a
statistical manner through the frequency distribution of those systems with
$\dot{f} < 0$. This may tell us the nature of the progenitors that survive
the onset of mass transfer. 

\subsection{Spin--orbit coupling and AM~CVn numbers}
The degree to which detached DWDs survive mass transfer to live on as
gravitational wave sources is highly dependent upon the degree to which
angular momentum accreted onto the more massive component is fed back into the
orbit \cite{Marsh:2004}. Some theoretical studies have indicated that this
coupling is strong and stabilising \cite{Racine:2007,Motl:2007}, while others
suggest that much of the angular momentum can be fed back even in the absence
of tidal coupling between the stars \cite{Sepinsky2010}. However the best
observational calibration of the space density of AM~CVn stars gives a space
density around a factor of 10 lower than previously assumed, and around 250
times lower than that of the detached systems, suggesting perhaps that in fact
many detached DWDs do not survive mass transfer \cite{Roelofs:2007}. This
issue is not settled but is another that \emph{LISA} is ideally suited to
answer: the ratio of systems with $\dot{f} > 0$ to those with $\dot{f} < 0$ as
a function of $f$ will be of great interest for addressing this question.

Although one can probably assume that a system with $\dot{f} < 0$ is an AM~CVn
star, the reverse is not true, i.e. $\dot{f} > 0$ does not imply a DWD, even
if we expect this to be the case more often than not. AM~CVn stars must pass
through an initial phase during which $\dot{f} > 0$. Indeed it is possible
that the two shortest period candidate AM~CVn stars, HM~Cnc and V407~Vul, are
in precisely this phase as both have decreasing orbital periods
\cite{Dantona:2006,Deloye:2007}. As first pointed out by
\citeasnoun{Webbink1998}, and further investigated by
\citeasnoun{Nelemans2004} and \citeasnoun{Stroeer2005}, the ``braking index''
$n = f \ddot{f} /\dot{f}^2$ is an interesting parameter in these cases. For
pure GWR-driven evolution, $n = 11/3$; we expect $n < 11/3$ during the turn-on
phase of AM~CVn stars, and during some phases $n < 0$. The second derivative
$\ddot{f}$ leads to a cubic dependence of binary phase on time, which places a
high value on extending \emph{LISA}'s lifetime for as long as possible:
without it we will not be able to distinguish detached DWDs from early-phase
AM~CVn stars except on a statistical basis, or perhaps through optical
follow-up of nearby systems.

The braking index has one further use. The standard $n = 11/3$ value for
detached DWDs treats the two stars as point masses, but as they approach the
onset of mass transfer we can expect the effects of tidal spin--orbit coupling
to become significant \cite{Willems2008}, in effect acting as an additional sink of orbital
angular momentum. Scaling as a high inverse power of the separation, tidal
losses will act to increase the value of $n$. \emph{LISA} detections of
systems with $n > 11/3$ may therefore provide a direct indication of the
significance of tidal coupling effects between white dwarfs.

\section{Combined EM \& GW observations}
A significant number of the DWDs that \emph{LISA} will see are potentially
detectable through optical observations. Several hundred with $V < 24$ are
predicted \cite{Nelemans2009}.  The bias towards short periods means that many
will eclipse \cite{Cooray:2004} (although the first, and at the moment only,
eclipsing detached DWD known, NLTT~11748, \citeasnoun{Steinfadt:NLTT}, has a
surprisingly long $5.6\,$ hour period). Eclipsing systems allow measurement of
the scaled radii, $R_1/a$ and $R_2/a$. Moreover, they permit precise optical
timing measurements which are quite capable of determining the conjunction
phases to within $<0.01$ cycles within a single night of observation. Optical
follow-up of such systems could add significantly to the numbers of systems
for which we know the first and second derivatives, $\dot{f}$ and $\ddot{f}$,
and hence the braking index $n$. The first challenge, as recognised by
\citeasnoun{Cooray:2004}, will be to locate them once \emph{LISA} has
signalled their presence, but with projects such as the \emph{LSST} underway,
this seems feasible, even given the large-by-optical-standards
\emph{LISA} error boxes. Photometric measurements alone have the capability to
determine the orbital inclination $i$ as well as the scaled radii. Using
mass-radius relations this could determine the mass ratio, which combined with
the chirp mass can lead to the two masses, giving a major insight into the
past evolution of the binaries. Spectroscopic observations will be difficult
given the faintness and short periods of most of the \emph{LISA}
targets. However, it should be noted that purely photometric observations may
well be able to return kinematic information entirely equivalent to
spectroscopy through Doppler beaming. This has already been detected in the
DWD eclipser, NLTT~11748, \cite{Shporer:2010}. This effect may even 
allow the identification of non-eclipsing optical counterparts using standard
phase-locked detection techniques. In the best cases there will result a
redundant set of constraints which will allow tests of white dwarf mass-radius
models. The result could be a bonanza for white dwarf astrophysics and provide
the best \emph{LISA} calibration sources.

\section{Conclusions}
Double white dwarfs are predicted to be the dominant source population at
\emph{LISA} frequencies. \emph{LISA}'s sensitivity to short orbital periods
will allow the best estimates of the merger rates of these stars, tidal
coupling of the two stars and answer questions about their evolution that are
hard to solve at the longer periods favoured by electromagnetic
observations. The dual combination of the GW and EM observations will be a
powerful tool for probing white dwarf astrophysics.

\section*{Acknowledgements}
I am indebted to Danny Steeghs, Boris G\"ansicke, Gijs Nelemans and Lars
Bildsten for discussions on these objects over the years and thank the
referees (Gijs Nelemans and one anonymous) for useful comments. The work was
carried out with financial support of the UK's Science and Technology
Facilities Council.

\section*{References}
\bibliography{marsh}{}
\bibliographystyle{jphysicsB}

\end{document}